# A mechanical experimental setup to simulate vocal folds vibrations. Preliminary results

**Nicolas Ruty**
**Annemie Van Hirtum**
**Xavier Pelorson**
*ICP, UMR 5009 CNRS/INPG/Université Stendhal Grenoble, France*

**Ines Lopez**
**Avraham Hirschberg**
*TU/e, Eindhoven, The Netherlands*

This paper contributes to the understanding of vocal folds oscillation during phonation. In order to test theoretical models of phonation, a new experimental set-up using a deformable vocal folds replica is presented. The replica is shown to be able to produce self sustained oscillations under controlled experimental conditions. Therefore different parameters, such as those related to elasticity, to acoustical coupling or to the subglottal pressure can be quantitatively studied. In this work we focused on the oscillation fundamental frequency and the upstream pressure in order to start (on-set threshold) either end (off-set threshold) oscillations in presence of a downstream acoustical resonator. As an example, it is shown how this data can be used in order to test the theoretical predictions of a simple one-mass model.

## 1. Introduction

The interaction of expiratory airflow with the vocal folds tissues is known to be the primary source of human voiced sound production (Titze, 1977; Fant, 1980). The air-flow through the larynx induces instability of the vocal folds. The resulting vocal fold vibrations modulate the airflow giving rise to a periodic sequence of pressure pulses which propagates through the vocal tract and is radiated as voiced sound. Modelling of the ongoing fluid-structure interaction and the vocal fold oscillations is important in the understanding of phonation (Fant, 1982a; Fant, 1982b; Ikeda et al., 2001), the synthesizing of voiced sound (Ishizaka et al., 1972; Koizumi, 1987; Story, 1995), and the study of voice disorders. Physical modelling of the vocal folds and the 3D fluid-structure interaction between the living tissues and the airflow has a long and rich history. Exact solutions for the flow through the glottis, as well as the flow-induced vocal fold deformation, are impossible to derive analytically. Full numerical simulation is still limited to over-simplified glottal configurations or flow conditions and is therefore not of practical use for many applications such as speech synthesis. Simplified models, such as distributed models (Ishizaka et al, 1972; Liljencrants, 1991; Kob, 2001), sometimes described as "caricatures", are therefore of interest for many practical applications. Concerning these simple models, two major problems can be formulated:

(1) how accurate are these simplified theories ?
(2) how can they be related to human physiology ?

One way to answer the first question (1) is to test the theoretical models against physical measurements performed on mechanical models of the vocal folds. Since the pioneer work of van den Berg (1957) many research groups in the world have developed mechanical replicas for this purpose and with an increasing complexity (and expected realism) Scherer et al., 1983; Pelorson et al., 1996; Gauffin, 1988. At present time, the most realistic set-ups are certainly those simulating the vibrations of the vocal folds (Barney, 1995; Zhao et al., 2002; Zhang et al., 2002). In this paper we present a new set-up capable of generating self-sustained oscillations of an elastic replica of the vocal folds. This mechanical model is interesting because it allows to generate and to measure unsteady flows at a rate comparable to those encountered during speech (50Hz $<F0<$1kHz). Of great interest for many applications, parameters associated to the onset and offset of vibration can also be measured. In addition, since the replica is producing sound, the acoustics can also be addressed using this setup as it will be shown in this paper.

Lastly, we will show how the second problem (2) can be addressed and tested using this experimental setup. It will be shown how the mechanical parameters of a simple one-mass model can be determined by direct observations performed on the replica.

## 2. Experimental setup

### 2.1. Description

The experimental set up is illustrated in figure 1. Up-scaled by a factor 3, it represents the human phonatory system composed of several elements. A pressure reservoir (approximate volume 0.75m³) represents the lungs and forces an airflow through a deformable vocal fold replica. Airflow is supplied by a compressor, which controls the pressure in the reservoir up to 50000Pa. The air pressure at the entrance of the reservoir is controlled thanks to a Norgren precision pressure regulator. The walls of the reservoir are covered with absorbing foam, in order to reduce its acoustical resonance.

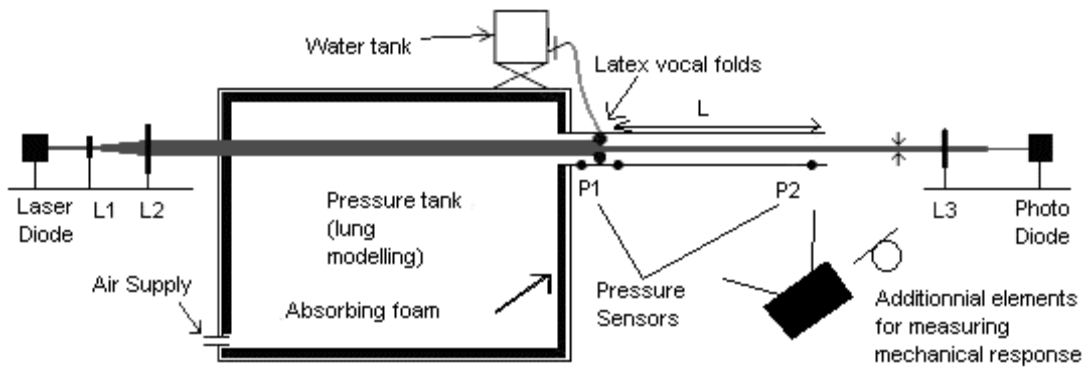

**Figure 1**: Schematic overview of the experimental set up used to produce self sustained oscillations and to detect the upstream pressure threshold necessary to obtain these oscillations. The water tank is used to control the internal pressure of the vocal folds replica.

The replica shown in Figure 2, is made of two latex tubes (Piercan Ltd) of 1.1cm diameter and 0.2mm latex thickness. Precision is on 10 percent on thickness and 1 percent on diameter. The tubes are mounted on a metal cylinder of 1cm diameter of which half of the diameter has been removed over a length of 3cm.

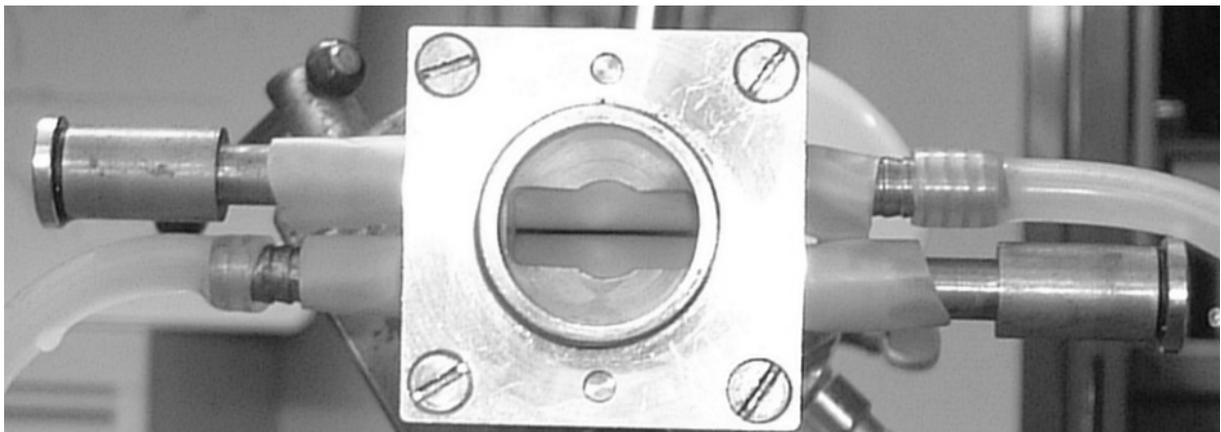

**Figure 2:** Vocal fold replica, made of two latex tubes, filled with water.

A central duct of 2mm diameter along the axis of the cylinder allows to fill the replica with water of which pressure (henceforth called the internal pressure Pc) is controlled by a water column. Changing the internal pressure Pc involves different types of changes. Firstly, because increasing the internal pressure has the consequence to fill the replica with more water, the mass of the replica increases. Secondly, the initial deformation of the replica is changed too. When internal pressure Pc is low (e.g. Pc=4000Pa), the replica remains well rounded as suggested in figure 3 whereas when the internal pressure is increased, the vocal fold replica are contacting which involves a more complex glottal geometry. Lastly, of course, increasing the internal pressure decreases the elasticity of the vocal fold replica. This vocal fold replica is connected with the reservoir directly.

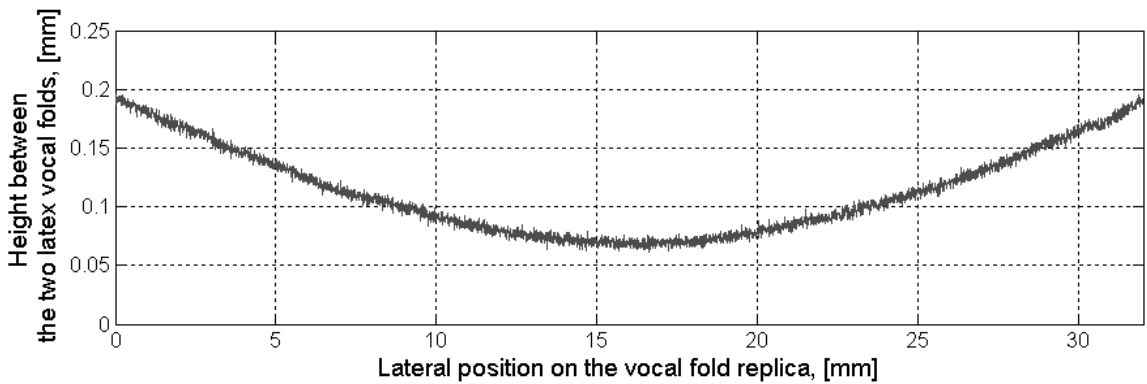

**Figure 3:** Frontal view of the deformable vocal fold replica, realised displacing laterally the laser beam and the photo sensor. Internal pressure Pc = 4000Pa

A downstream resonance pipe, simulating a vocal tract, of varying length L is also connected downstream of the glottal replica. Its section is constant, of 2.5cm diameter. Two Kulite pressure sensors XCS-093-0.35-Bar-G (P1, P2), supplied by a Labor-Netzgerät power supply EA-3005S, enable us to measure upstream pressure and downstream pressure respectively. The pressure sensors were calibrated against a watermeter with typical accuracy of +/- 5Pa.

Thanks to an optical laser system, the replica opening h, i.e. the vertical 1D movement of the deformable replica, can be recorded. This optical system consists in a laser diode (supplied by a P. Fontaine Dc amplifier FTN2515) which laser beam is increased by two convergent lenses (L1, f=50mm; L2, f=100mm), at a distance of 125mm. Then, the laser beam goes through the reservoir, and further through the vocal folds replica. When oscillating, the glottal replica is therefore interrupting the laser beam. The corresponding intensity fluctuations are recorded by a photo sensor (supplied by a Solartron DC Power Supply).

## *2.2. Data acquisition*

This system is linked to an acquisition chain which enables us to detect the presence of oscillation as well as the oscillation frequency from the registered changes in replica aperture.

This data acquisition system consists in a filtering preamplifier (National Instruments SXCI-1121) in which all sensor signals are filtered and amplified between +10V and -10V. The outputs of this filter are plugged onto a National Instruments BNC2080 card linked to an acquisition card National Instruments PCI-MIO-16XE. Lastly the acquired data are processed in Labview7.

## *2.3. Calibration*

Pressure sensors, optical system and measurement microphone have to be calibrated.

Pressure sensors have a linear response. We have to determine their gain. Airflow is forced in a rigid tube. Pressure is simultaneously measured with a Kimo pressure gauge and with a pressure sensor. For each value of pressure ranging between 50 and 1000Pa, the real pressure values measured from the pressure gauge are related to the electrical voltage measured by the pressure sensors. From the linear experimental curve the gain of each pressure sensor is estimated.

The optical system is used to measure the vocal folds replica opening. The intensity of the light receives by the photo sensors have to be related to the opening of the obstacle the laser beam crosses. For the calibration the replica is replaced by a rigid rectangular aperture of a height ranging between 0.05mm and 3mm. Each value of the aperture is related to the voltage measured by the photo sensors.

Lastly the Bruël and Kjaer measurement microphone is calibrated with a Bruël and Kjaer Sound Level Calibrator Type 4230 which produce a pure sinusoidal audio signal of 1000Hz frequency and 94dB amplitude.

## 3. Results

### *3.1. Oscillation threshold*

The set-up illustrated in Figure 1 and 2 is able to produce self-sustained oscillations. In the experiments detailed here, a downstream pipe of length L=49cm was used. The internal pressure of the vocal folds was ranging between 3500 and 6500Pa. In order to get this range of variation, the height of the water column that fills the replica was ranging between 35cm and 65cm.

For a given internal pressure, the upstream pressure is increased until oscillations appear. Oscillations are detected by spectral analysis of the pressure signal recorded by the sensor P2. The presence of oscillations is validated as soon as the spectrum is composed of at least a fundamental frequency and two superior order harmonics. Hence we obtain the mechanical experimental phonation on-set threshold.

When stable oscillations are achieved, the fundamental frequency of oscillation is recorded and the upstream pressure is decreased. The same procedure is followed to retrieve the transition from self-sustained oscillation to no-oscillation state. The upper upstream pressure for which oscillations disappear corresponds to the phonation off-set threshold.

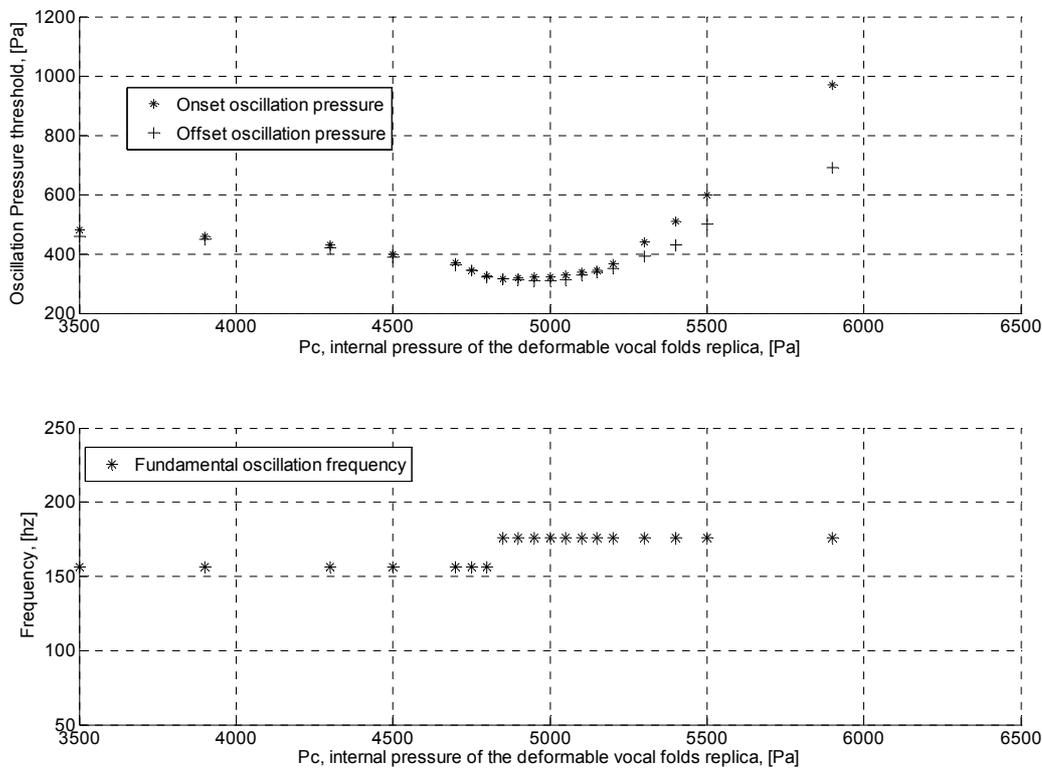

**Figure 4:** (a), Upstream pressure oscillation threshold, as a function of the internal deformable replica pressure. (b), fundamental frequency of the oscillation.

Figure 4 shows an example of the results obtained using this procedure. A lot of observations can be made. Firstly, the range of values we obtain for the threshold is comparable to the human phonation threshold, with an average value of 400Pa (Baken, 1987). Secondly, we notice that a hysteresis phenomenon is present when we compare the on-set and the off-set pressure as noted by Lucero, 1999. Furthermore a minimum threshold pressure is reached at Pc=5000Pa. That confirms further studies considering the oscillation threshold (Chan et al, 1997; Titze et al., 1995). More precisely, we observed that this minimum is reached when the vocal folds replica is almost closed at rest position (Figure 5). This behaviour can be related to the optimum configuration for the ease of phonation discussed by (Lucero, 1998).

### *3.2. Oscillation frequency*

As illustrated in Figure 4, the fundamental frequency of oscillations observed experimentally is close to the downstream pipe acoustical resonance. Indeed, the resonance frequency of the pipe is equal to 173.5Hz and what we observe in Figure 4 is fundamental frequency ranging between 155Hz and 180Hz. Fundamental frequencies seem to be constrained between a short range of values. We can note that further measurements have to be done. Indeed, we observe only two values for fundamental frequencies. It is in fact not the case, but for real time measurements, we have to reduce the FFT scale and loose in precision for the measurement of fundamental frequency. What we obtain is an order of range for the fundamental frequency of the oscillations (with a precision of 10Hz approximately).
Then, concerning this range, we observe that such values are between male and female phonation fundamental frequencies.

### *3.3. Influence of initial parameters*

In this section, we examine the influence of the internal pressure, $P_c$ as a control parameter for the deformable vocal folds replica. Of particular interest is its effect on the equilibrium position of the vocal folds and on their mechanical characteristics. The consequences of these changes can be observed in terms of oscillation thresholds and oscillation frequencies since changing the internal pressure, $P_c$ affects the mechanical characteristics. Therefore the equilibrium aperture of the replica in absence of upstream and downstream resonators as well as the mechanical response of the deformable replica is measured. Measurement results are detailed below.

### *3.3.1. Equilibrium positions*

The procedure used to determine the equilibrium positions is as follows. For each given internal pressure, $P_c$ (ranging between 1500Pa and 6500Pa), the upstream pressure is increased from 50Pa to 1000Pa by steps of 50Pa. Thanks to the optical laser system, the aperture between the two latex tubes can be measured. The results are presented in Figure 5.

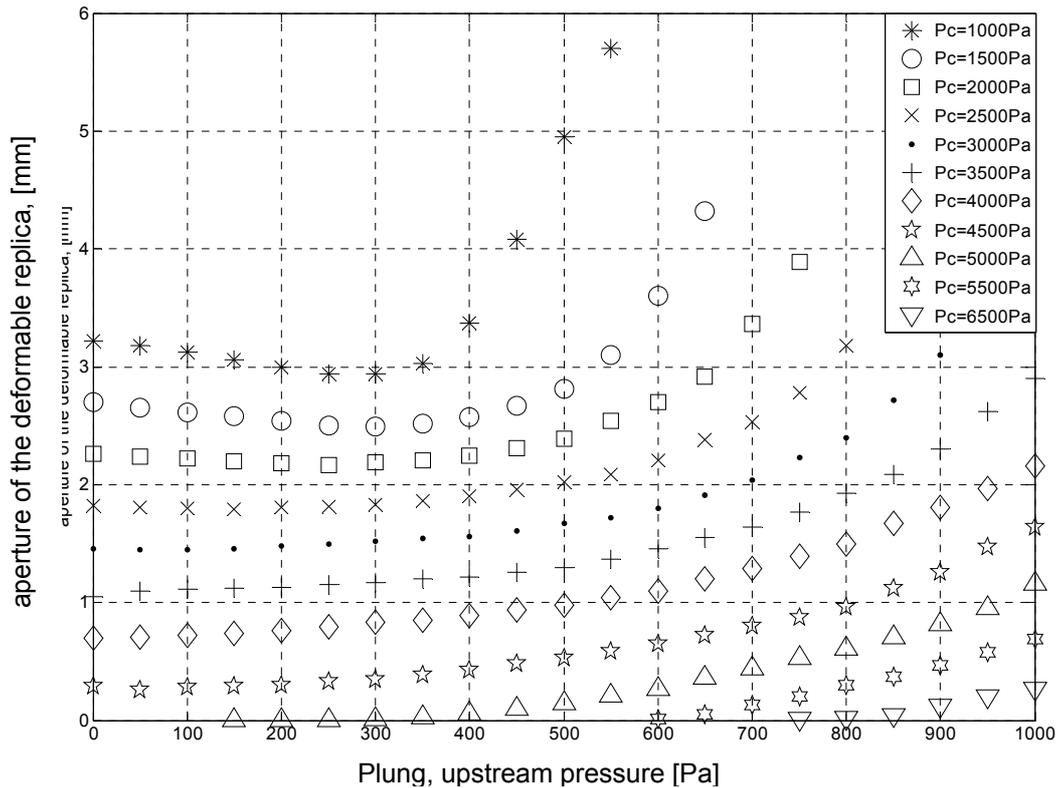

**Figure 5:** Equilibrium position of the deformable vocal fold replica measured as a function of the upstream pressure (Plung), for different values of the internal pressure (Pc) in the deformable replica

From this figure some interesting behaviour can be observed. For low internal pressures, namely lower than 3000Pa, increasing the upstream pressure at first decreases the replicas aperture until a minimum opening is reached; further pressure raise increases the equilibrium aperture. When the internal pressure Pc is higher than 3000Pa, increasing the upstream pressure results in forcing the vocal folds replica to open.

Concerning the position at rest, without upstream pressure, we observe that increasing internal pressure involves a decrease of the aperture until the tubes are in contact for internal pressures up to 5000Pa. 5000Pa appears to be a critical value for the internal pressure. This corresponds to the frontier between an opened rest position and a condition where the vocal folds are contacting at rest. It also corresponds to the point where oscillation pressure threshold is minimal, as described in section 3.1.

We can conclude from this measurement that there is a strong correlation between the oscillations threshold and the equilibrium position.

*3.3.2. Mechanical response*

In addition to that, we use a measurement microphone Bruël and Kjaer type 4192 with its preamplification unit Bruël and Kjaer and a loud speaker driven by an Onkyo Integra Stereo Amp.

This second measurement consists in the determination of the mechanical response of the deformable vocal folds replica from which we extract the resonance frequency and a quality factor.

The experimental protocol is the following. In addition to the experimental set-up depicted in Figure 1, a loud speaker and a Bruël and Kjaer (type 4192) measurement microphone were added in the vicinity of the vocal folds replica. Using the loudspeaker the replica was excited using a sinusoidal signal with a frequency varying between 100Hz and 400Hz. The varying maximal opening h is captured during the vertical movements of the replica by way of the photo-diode through the acquisition chain. Hence, after filtering, the needed parameters (resonance frequencies, quality factor and equilibrium position) are derived. This filtering consists in a deconvolution of the displacement, by the microphone signal.

An example of mechanical response measured for internal pressures Pc varying from 1500 to 6500Pa is presented in Figure 6. Resonance frequencies are detected if a peak is more important (higher than 10dB) in terms of its amplitude, than the mean signal. We focus our attention on the peak close to the one observed during the oscillations (see 2.1). The quality factor is calculated using the following formula:

$$Q = \frac{f_{res}}{\Delta f_{-3dB}}, \quad (1)$$

where $f_{res}$ is the resonance frequency and $\Delta f_{-3dB}$ is the associated 3dB bandwidth.

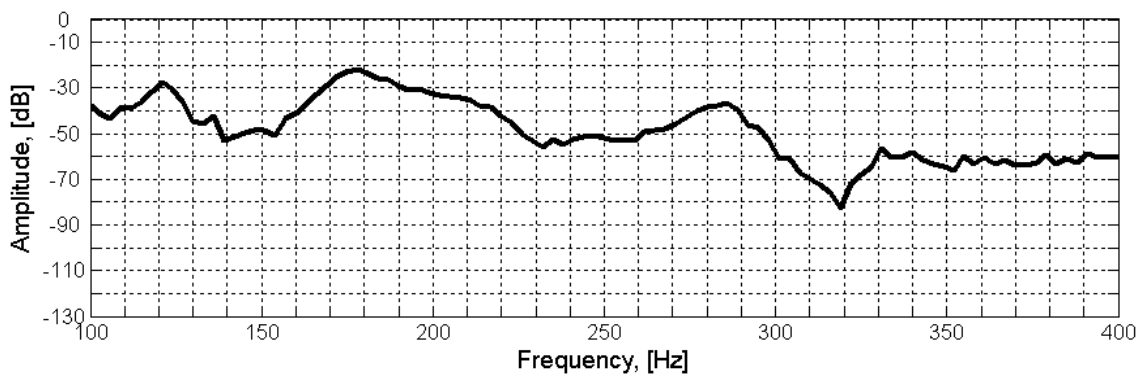

**Figure 6:** Mechanical response of the deformable vocal folds replica, for an internal pressure of 6000 Pa.

The experimental parameters obtained from these experiments are tabulated in Table 1. We observe that resonance frequency is close to the one measured when the replica self-oscillates. However, increasing internal pressure involves an increase in resonance frequency.

Table 1: Recapitulative table of the different experimental parameter values used for the numerical simulation.

| Internal Pressure Pc(Pa) | 3500 | 4000 | 4500 | 5000 | 5500 | 6000 | 6500 |
|---|---|---|---|---|---|---|---|
| Resonance frequency (Hz) | 123 | 124 | 130 | 140 | 166 | 169 | 175 |
| Quality Factor | 8.2 | 8.267 | 8.3 | 10 | 5.5 | 14.08 | 17.5 |

This effect is to be related again to a variation of the stiffness of the replica. High internal pressure involves a higher vocal folds replica stiffness and thus higher resonance frequency as observed in real life on human speakers.

Another observation is that the resonance frequencies are varying between 123Hz and 175Hz while when self-sustained oscillations where present frequencies were observed in a smaller range (between 155Hz and 180Hz). This tends to illustrate the effect of the downstream resonator on the oscillation frequencies of the replica.

In the absence of extensive data concerning the effect of the subglottal system on human phonation, the experimental study of the influence of an upstream pipe in the mechanical setup is currently omitted since it is difficult to conclude about the relevance.

### 4. Using the measured data in physical models

As an example of application we present in the following an attempt to use the set-up for testing a physical model of human phonation. For the sake of simplicity, the model chosen will be based on a simple one mass model of the vocal folds. This model is depicted in figure 7.

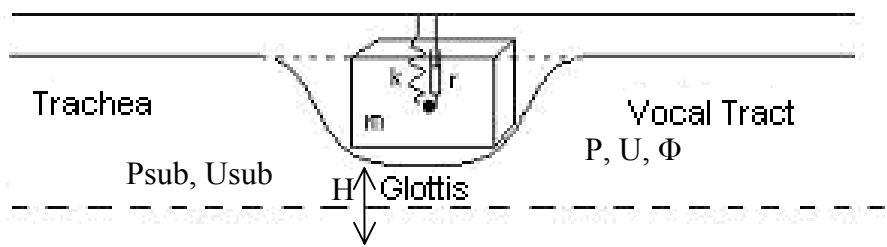

**Figure 7:** One spring mass model of the vocal folds, with k, r and m respectively the stiffness, the damping and the mass of a vocal fold. H is the displacement, between the two vocal folds, Psub and Usub respectively the upstream pressure and flow velocity. Φ is the volume flow, P and U respectively the downstream pressure and flow velocity.

Assuming small variations around the equilibrium position, this system, coupled with a downstream resonator (a uniform section pipe), can be described with the following set of equations:

$$\frac{d^2 h}{dt^2} + \frac{\omega_l}{Q_l}\frac{dh}{dt} + \omega_l^2 h = -\frac{1}{\mu} p \tag{1}$$

$$\phi = b\left(-\frac{\bar{h}\, p}{\rho\, \bar{u}} + \bar{u}\, \bar{h}\right), \quad \bar{u} = \sqrt{\frac{2\,\bar{p}_{sub}}{\rho}} \tag{2}$$

$$\frac{d^2\psi(t)}{dt^2} + \frac{\omega_a}{Q_a}\frac{d\psi(t)}{dt} + \omega_a^2\psi = \frac{Z_a\omega_a}{Q_a}\phi \tag{3}$$

(1) is the mechanical equation describing the elasticity behaviour of the vocal folds. h is the variation of the glottal opening around the equilibrium position $\bar{h}$. p is the variation of transglottal pressure. $\omega_l$ is the resonance pulsation. $\omega_l = \sqrt{k/m}$, k is the stiffness of the spring, m the mass. $Q_l$ is the quality factor associated to this resonance. $Q_l = \omega_l / r$, r is the damping of the one mass model.

(2) is the fluid mechanical equation describing the behaviour of the airflow through the glottis. Φ is the volume flow. b is the glottal width, ρ is air density, $\bar{u}$ is the mean airflow velocity, $\bar{p}_{sub}$ is the mean subglottal (or upstream pressure)

(3) is the acoustical equation describing the propagation of acoustic waves inside the vocal tract, approximated here as a uniform pipe. $p = d\psi/dt$, where $\psi$ is the acoustical eigenfunction. $\omega_a$, $Q_a$, $Z_a$ are the resonance pulsation, the quality factor and the static impedance of the pipe respectively.

While geometrical or acoustical parameters in the above equations can be determined directly, the same cannot be said about the mechanical characteristics in equation (1). Therefore, for each internal pressure, Pc, the values attributed to the parameters $Q_l$ and $\omega_l$ were obtained in order to fit the experimental mechanical response (see Table 1).

Then for subglottal pressure $\bar{p}_{sub}$ ranging between 20 and 1000 Pa a linear stability analysis method, as proposed by Cullen et al. (2000), was performed.

Hence the results of simulation can be compared to the experimental ones as shown in figure 8.

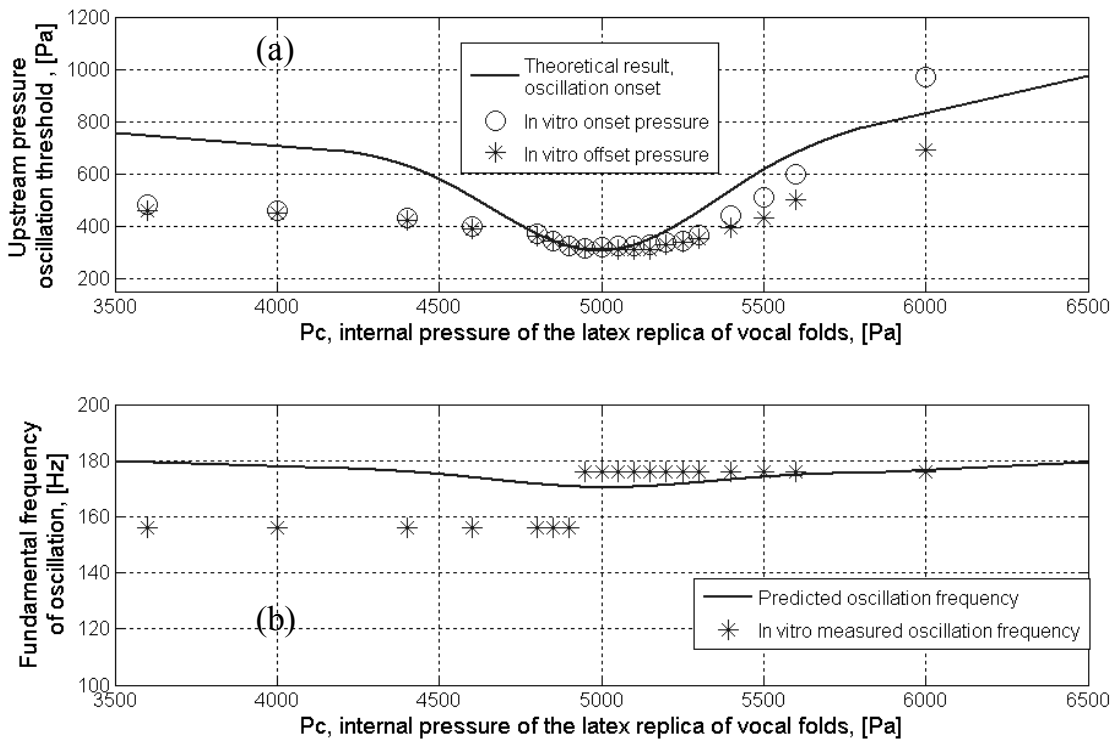

**Figure 8:** (a), Upstream pressure oscillation threshold, comparison of experimental results and theoretical prediction. (b), Comparison of experimental results and prediction for the fundamental frequency of the oscillation.

On the same graph (figure 8a), theoretical and experimental threshold are presented showing that even such a crude model can explain qualitatively the experimental results. Moreover, fundamental frequencies are also theoretically predicted and compared to the experimentally measured ones as shown in figure 8b. Once again, theory appears to be close to experiments. The same phenomena are observed on the two curves, i.e. the presence of a minimum threshold for an internal pressure of 5000Pa and a fundamental oscillation frequency close to the downstream pipe (vocal tract) resonance frequency.

## 5. Conclusion

In this paper we have presented a new experimental set-up designed in order to test physical models for human voice production. This set-up presents a self-oscillating vocal folds replica made of latex filled with water which is coupled to acoustical resonators (representing the subglottal or the supraglottal tract). This allows to measure aerodynamical (the coupling between the elastic structure and the flow) and aeroacoustical (the coupling between the acoustic field and the flow) phenomena at the same time, which, to the best of our knowledge, has never been done using a mechanical replica. Further, compared with existing set-ups this one has the advantage to generate the self-sustained oscillations of a replica whose mechanical characteristics (controlled using the internal pressure, $P_c$) can be controlled quantitatively. This is particularly important when willing to compare to low order mechanical models such as the one- or the two-mass model.

The first results obtained with this replica tend to be encouraging. Oscillation threshold pressures were found quite comparable to those observed in human speech and with comparable frequencies, at least for healthy male speakers. Even more interesting is the experimental evidence for a minimum threshold pressure which was already theoretically predicted by Lucero (1998).

Of course, the greatest care should be taken in the interpretation of the measured data. This set-up has been built in order to test the theoretical models, as shown for the one-mass model example here, rather than to mimic reality.

Further study is therefore ongoing, testing systematically different approximations for the mechanical behaviour of the vocal folds as well as the effect of acoustical coupling. The set-up itself is also under study, in particular in order to generate a broader quantity of configurations such as the presence of asymmetrical vocal folds, high frequency fundamental frequency of oscillation, etc …

### *Acknowledgements*

This work has been supported by the Region Rhône-Alpes (project Emmergence) a PhD grant from the French Minister of Research and Education. We would like to acknowledge the skills of Pierre Chardon who helped us to design and to build the set-up. Lastly, we wish to thank Susanne Fuchs and an anonymous reviewer for useful remarks.